\documentclass[a4paper,fleqn,usenatbib]{mnras}

\usepackage{newtxtext,newtxmath}

\usepackage[T1]{fontenc}
\usepackage{ae,aecompl}

\usepackage{graphicx}	
\usepackage{amsmath}	

\title[Realistic simulations find significant entrainment]{Realistic 3D hydrodynamics simulations find significant turbulent entrainment in massive stars}

\author[Rizzuti et al.]{F. Rizzuti$^{1}$\thanks{E-mail: f.rizzuti@keele.ac.uk}, R. Hirschi$^{1,2}$, C. Georgy$^{3}$, W. D. Arnett$^{4}$, C. Meakin$^{5}$ and A. StJ. Murphy$^{6}$
\\
$^{1}$Astrophysics Group, Lennard-Jones Laboratories, Keele University, Keele ST5 5BG, UK\\
$^{2}$Kavli IPMU (WPI), University of Tokyo, 5-1-5 Kashiwanoha, Kashiwa 277-8583, Japan\\
$^{3}$Geneva Observatory, Geneva University, CH-1290 Sauverny, Switzerland\\
$^{4}$Steward Observatory, University of Arizona, 933 N. Cherry Avenue, Tucson AZ 85721, USA\\
$^{5}$Pasadena Consulting Group, 1075 N Mar Vista Ave, Pasadena, CA 91104 USA\\
$^{6}$School of Physics and Astronomy, University of Edinburgh, Edinburgh EH9 3FD, UK
}

\date{Accepted XXX. Received YYY; in original form ZZZ}

\pubyear{2022}

\begin{document}
\label{firstpage}
\pagerange{\pageref{firstpage}--\pageref{lastpage}}
\maketitle

\begin{abstract}
Our understanding of stellar structure and evolution coming from one-dimensional (1D) stellar models is limited by uncertainties related to multi-dimensional processes taking place in stellar interiors. 1D models, however, can now be tested and improved with the help of detailed three-dimensional (3D) hydrodynamics models, which can reproduce complex multi-dimensional processes over short timescales, thanks to the recent advances in computing resources. Among these processes, turbulent entrainment leading to mixing across convective boundaries is one of the least understood and most impactful. Here we present the results from a set of hydrodynamics simulations of the neon-burning shell in a massive star, and interpret them in the framework of the turbulent entrainment law from geophysics. Our simulations differ from previous studies in their unprecedented degree of realism in reproducing the stellar environment. Importantly, the strong entrainment found in the simulations highlights the major flaws of the current implementation of convective boundary mixing in 1D stellar models. This study therefore calls for major revisions of how convective boundaries are modelled in 1D, and in particular the implementation of entrainment in these models. This will have important implications for supernova theory, nucleosynthesis, neutron stars and black holes physics.

\end{abstract}




\begin{figure*}
\centering
\footnotesize
\includegraphics[trim={1.5cm 1cm 1.5cm 1cm},clip,width=0.5\textwidth]{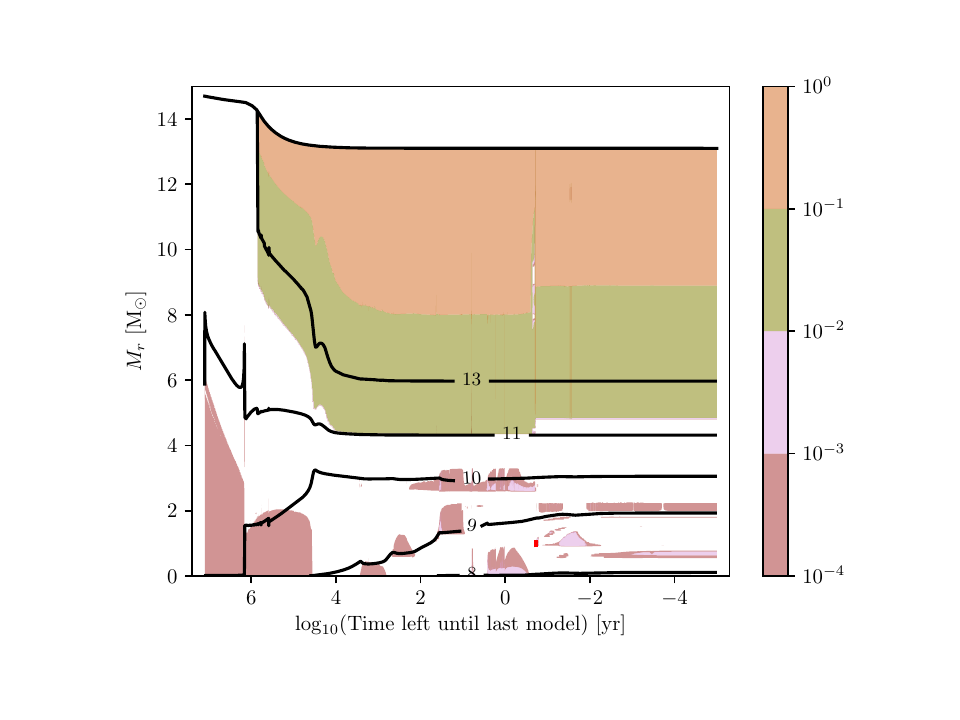}\includegraphics[trim={1.5cm 1cm 1.5cm 1cm},clip,width=0.5\textwidth]{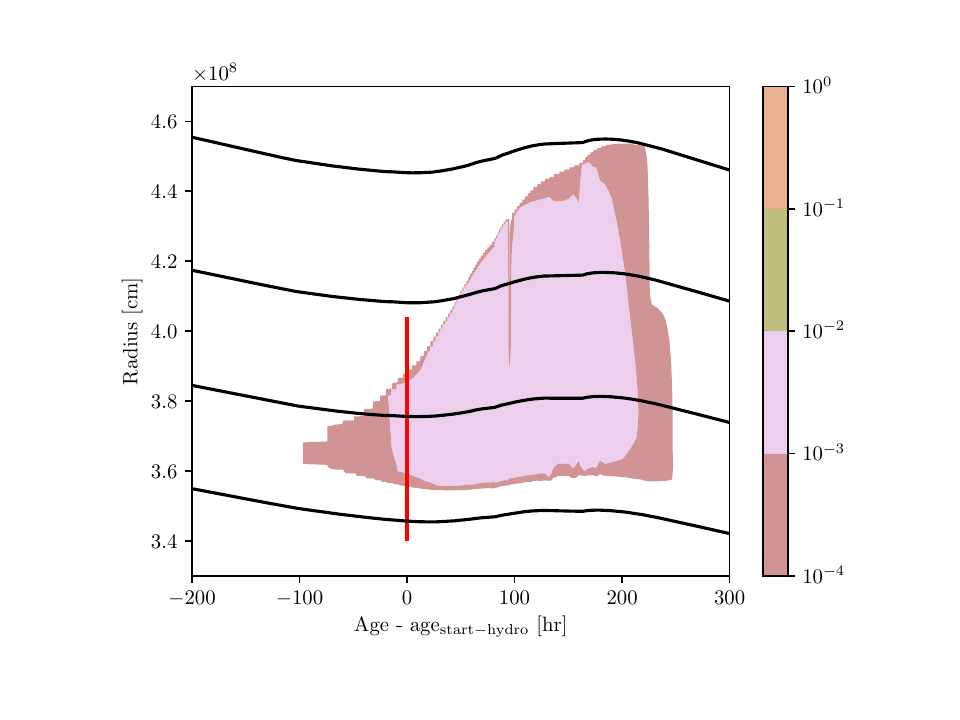}\put(-366.5,23.5){\color{red}\huge\boldmath$\uparrow$}
\caption{{\it Left}: Structure evolution diagram of the 15 M$_\odot$ 1D \texttt{GENEC} input stellar model as a function of the time left until the predicted collapse of the star (in years, log scale). The last model calculated is in between oxygen and silicon burning. The total mass and radial contours are drawn as black lines, with numbers indicating values of log$_{10}$(r) in cm. Shaded areas 
correspond to convective regions, the colour indicating the value of the Mach number. The vertical red bar around log$_{10}$(time left in years) $\sim -1$ and a mass coordinate of $\sim 1$ M$_\odot$ represents the domain simulated in 3D, and the time at which the 3D simulations start. {\it Right:}  Zoom-in around the model used as initial conditions for the 3D hydrodynamics simulations. The horizontal axis is the time relative to the start of the 3D simulations (age$_{\text{start-hydro}}$). The vertical axis is the radius in units of 10$^8$\,cm. Isomass contours for $M_r=0.9,1.0,1.1,1.2$\,M$_\odot$ are shown by black lines. These lines show that the shell studied does not undergo any significant contraction or expansion during the Ne-burning phase. The vertical red bar indicates the start time
and radial extent of the hydrodynamics 3D simulation. The physical time of the simulations is one hour or less, still much shorter than the time-scale of this plot. The shaded areas correspond to convective regions, the colour indicating the value of the Mach number.}\label{fig:kip}
\end{figure*}

\section{Introduction}
One-dimensional (1D) stellar evolution models \citep[e.g.][]{2002ApJ...567..532H,2011ApJS..192....3P,2012A&A...537A.146E} are powerful tools employed to study the structure, evolution and fate of stars over long timescales, often from their birth to their death. They can be used to probe different aspects of astrophysics, and they are the key to tackle many open problems related to gravitational waves \citep[e.g.\ black hole mass gap, ][]{2021ApJ...912L..31W} and supernova explosions (e.g. role of progenitor compactness, \citealt{2011ApJ...730...70O, 2016ApJ...818..124E}, and the red supergiant problem, \citealt{2009ARA&A..47...63S}) due to their dependence on the progenitor structure and properties. The most important assumption implicit in 1D models is that of spherical symmetry. This assumption enables 1D simulations to cover large spatial and temporal extents, but inevitably results in loss of information coming from deviations from spherical symmetry. For this reason, complex physical phenomena like turbulent convection can be included in the models only by means of prescriptions, which require parameters usually calibrated to observations. The price to pay is the omission of terms such as the turbulent kinetic energy flux, and large uncertainties related to convection, both of which limit the predictive power of stellar models \citep[see][]{2020MNRAS.496.1967K}. Convective boundary mixing (CBM), a physical phenomenon which takes place near the edges of convective zones, is one of the most challenging aspects of convection to include in 1D models. The two most commonly used implementations for CBM are penetrative overshooting \citep{1991A&A...252..179Z} and exponentially decaying diffusion \citep{1996A&A...313..497F}. The former provides an extension of the convective zone by a fraction of the pressure scale height \citep[e.g.][]{2012A&A...537A.146E}, while the latter extends the mixing of chemical composition beyond the boundary as a diffusive process \citep[e.g.][]{2000A&A...360..952H}. Both methods have been calibrated such that model outputs correctly match some observations \citep{2000A&A...360..952H, 1992A&AS...96..269S, 2011A&A...530A.115B, 2021MNRAS.503.4208S}. The same models, however, fail to reproduce observations outside their calibration range. One major example is the struggle of stellar models in reproducing the width of the main sequence for different mass ranges \citep{2014A&A...570L..13C}.\\
In recent decades, asteroseismic measurements have brought new light on this topic by both pointing out the weaknesses of current stellar models and providing guidance for new prescriptions for convection \citep[see][]{2019ARA&A..57...35A}. Multi-dimensional hydrodynamics simulations of stellar environments can help guide 1D model prescriptions through their ability to properly reproduce complex physical processes, including CBM, without assuming any prescription. This is especially needed for the late burning phases in massive stars, during which surface observables are detached from the interior evolution. The turbulent fluid motion, realistically simulated in great detail, can explain and predict the flow behaviour in stellar environments. Leading examples of using 3D simulations in stellar astrophysics include nova models \citep{1998ApJ...494..680J}, supernova models \citep{2015ApJ...807L..31L,2016Janka, 2021Natur.589...29B}, and the atmosphere of the Sun \citep{2009ARA&A..47..481A}. Unfortunately, the computing cost required to run these multi-D simulations is very high, so at present it is not yet possible to study the long-term evolution of stars in more than one dimension. Instead, results coming from short, localized multi-D simulations can be used to constrain and improve the simplified assumptions currently employed in 1D. Due to the computing cost, past multi-D simulations of convective zones had several limitations. One of them is the necessity of boosting the driving luminosity by large factors ($\geq$1000) to reduce the computing time \citep[see][]{2017MNRAS.471..279C,2021A&A...653A..55H}, which introduces important differences from the 1D input model and changes its equilibrium state. Furthermore, most stellar hydrodynamics codes do not include an explicit nuclear network to drive convection, but instead rely on fixed heating profiles \citep[e.g.][]{2017MNRAS.465.2991J}. Finally, another important limitation arises from simplified or unusual initial conditions, which may lead to entrainment of a different nuclear fuel \citep[e.g.][]{2007ApJ...667..448M} and invariably make the interpretation of results more difficult.
\\We organize the paper as follows: in Section 2, we describe the initial conditions and general setup of the hydrodynamics simulations. In Section 3, we present and analyse results from the set of simulations. Finally, we discuss the results and draw some conclusions in Section 4.

\begin{figure*}
\centering
\footnotesize
\includegraphics[trim={0cm 0cm 0cm 0cm},width=0.48\textwidth]{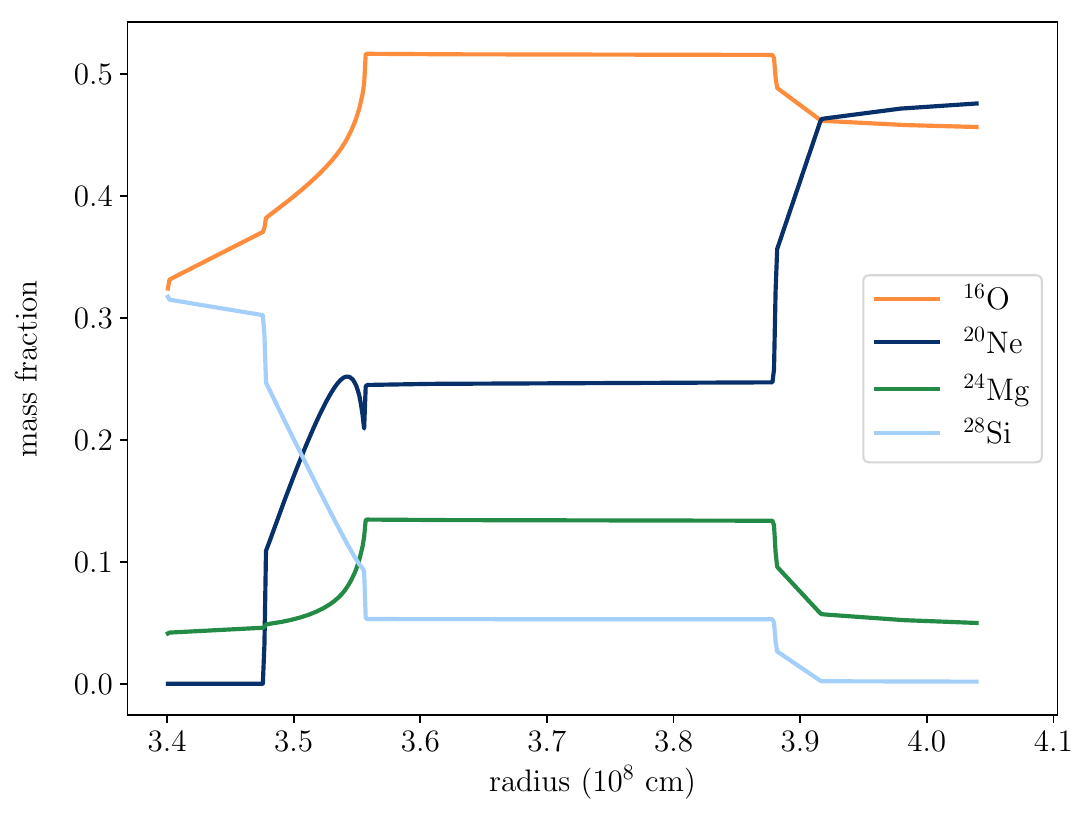}\hfill\includegraphics[trim={0cm 0cm 0cm 0cm},width=0.48\textwidth]{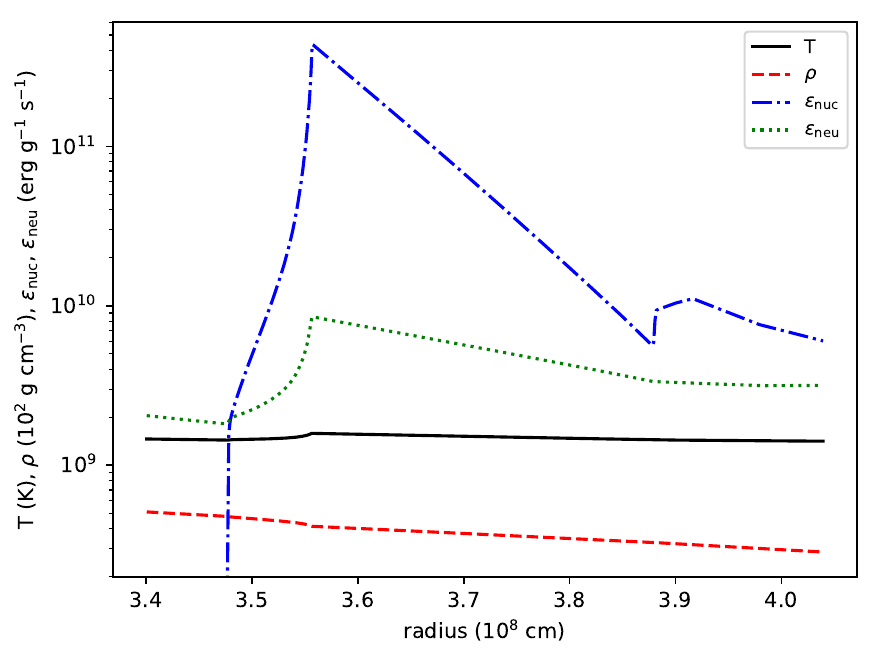}
\caption{Initial properties extracted from the 15 M$_\odot$ 1D \texttt{GENEC} input stellar model. {\it Left}: Abundance profiles for $^{16}$O, $^{20}$Ne, $^{24}$Mg and $^{28}$Si.
{\it Right}: Profiles for temperature ($T$, solid black line) density ($\rho$, red dashed line), nuclear energy generation rate ($\epsilon_{\textrm{nuc}}$) and neutrino energy loss rate ($|\epsilon_{\textrm{neu}}|$). These profiles show the temperature and most clearly the energy generation peak at the bottom of the convective zone. 
}\label{fig:abu_iniTrho}
\end{figure*}

\section{The simulations}
\subsection{Input 1D stellar evolution model}
In this study, we focus on modelling the neon-burning shell of a massive star. For this purpose, we computed a stellar evolution model for a 15 M$_\odot$ star at solar metallicity ($Z=0.014$) from the zero-age main sequence up to the start of silicon burning with \texttt{GENEC} \citep{2008Ap&SS.316...43E}. Standard physical ingredients described in \citet{2012A&A...537A.146E} were used. In particular, concerning convection, the Schwarzschild criterion is used and penetrative overshoot is included for core hydrogen and core helium burning phases only, with an overshooting distance $\ell_{\text{over}}=0.1\,H_P$. From carbon burning onwards, an $\alpha$-chain network is used (see \citealt{2004A&A...425..649H} for details). The evolution of the structure of the model is presented in Fig.\,\ref{fig:kip} ({\it left}). A zoom-in of the neon-burning shell used as initial conditions for the 3D simulations is presented in Fig.\,\ref{fig:kip} ({\it right}). 
\\Compared to other convective episodes, neon-burning convective zones are relatively small and short-lived with typical spatial extents of the order of $10^8$-$10^9$\,cm and temporal extents of weeks to months. These timescales, however, are still much longer than the duration of 3D hydrodynamics simulations (see Table\,\ref{tab:1}).
In time, neon-shell burning follows neon and oxygen core burning phases, whereas in space (or mass) neon-shell burning follows helium and carbon burning. The succession of burning stages leads to the abundance profiles in and around the neon shell used as initial conditions in this study (see Fig.\,\ref{fig:abu_iniTrho}, {\it left}). Above the convective zone (radius $r>3.88\times 10^8\,$cm), the composition is the result of helium burning producing mostly $^{12}$C and $^{16}$O, and then carbon burning converting $^{12}$C into $^{20}$Ne, leaving $^{16}$O and $^{20}$Ne as the most abundant nuclear species. Inside the convective neon-burning shell ($3.56<r<3.88\times 10^8\,$cm), easily recognised from the flat composition profiles, $^{20}$Ne is photo-disintegrated into $^{16}$O. These photo-disintegrations also release $\alpha$ particles that can capture on $^{20}$Ne nuclei to produce $^{24}$Mg. Some of the $^{24}$Mg produced can also capture $\alpha$ particles to produce $^{28}$Si.
Below the convection zone ($r<3.56\times 10^8\,$cm), radiative neon and oxygen burning have already partially taken place and converted oxygen, neon and magnesium into nuclides like silicon and sulphur. These multiple burning episodes explain why the $^{16}$O and $^{24}$Mg abundances are higher in the convective region than both below and above it. The $^{28}$Si abundance profile decreases outwards as expected. The $^{20}$Ne generally increases outwards. An exception to this monotonic outward increase for $^{20}$Ne  takes place just below the bottom of the convection zone ($r\sim 3.55\times 10^8\,$cm) and can be explained by the temperature inversion caused by neutrino energy losses being larger than nuclear energy generation in the core after core oxygen burning. This temperature inversion can be seen in the initial profiles shown in Fig.\,\ref{fig:abu_iniTrho} (solid black line in the {\it right} panel). Fig.\,\ref{fig:abu_iniTrho} ({\it right}) also shows the initial profiles for density, nuclear energy generation rate and neutrino energy loss rate. 
\\The following steps are undertaken to translate the 1D spherically-symmetric structure into the initial conditions for the plane-parallel 3D simulations presented in this study:
\begin{itemize}
\item all thermodynamic quantities ($T,\rho,P$), gravity and the chemical composition as a function of mass/radius are read from the 1D \texttt{GENEC} code output file called ``vfile'';
\item the Timmes equation of state \citep[EOS,][]{Timmes_1999} is used to calculate the entropy ($s$, missing in the \texttt{GENEC} output) and to recalculate consistently (\texttt{GENEC} uses a different EOS) the temperature as a function of density and pressure. The EOS is also used to calculate the relevant thermodynamic partial derivatives in order to determine d$\rho/$d$r$ in the final step; 
\item an analytical fit of the gravity profile is calculated in order to be used during the hydrostatic equilibrium equation (step below) and during the 3D simulations (gravity profile fixed in time):
\begin{equation}
  g(r\text{[cm]})= -\sum_{n=0}^{7} \,a_n\, r^n\, \text{cm s}^{-2},
\end{equation}
where {\footnotesize  $a_0-a_7=[1.38986906\times10^{14},\,-\, 2.61125476\times10^{6},\,2.10111788\times10^{-2},\,-\, 9.38605609\times10^{-11},\,2.51406418\times10^{-19},\,-\, 4.03771206\times10^{-28},\,3.60032241\times10^{-37},\,-\, 1.37497579\times10^{-46}]$}.
\item finally, the profiles of the thermodynamic quantities are integrated over a fine mesh of 5000 grid points starting from the bottom of the domain, combining the hydrostatic equilibrium and the gravity fit calculated above in order to integrate the density from 
    \begin{equation}
        \frac{\mathrm{d}\rho}{\mathrm{d}r} = \frac{\partial\rho}{\partial s}\frac{\mathrm{d}s}{\mathrm{d}r} +\frac{\partial\rho}{\partial P}\frac{\mathrm{d}P}{\mathrm{d}r} + \frac{\partial\rho}{\partial A}\frac{\mathrm{d}A}{\mathrm{d}r} + \frac{\partial\rho}{\partial Z}\frac{\mathrm{d}Z}{\mathrm{d}r}
    \end{equation}
where $A, Z$ are the mean atomic mass and number (also extracted from the \texttt{GENEC} output), and using for the second term the condition for hydrostatic equilibrium d$P/$d$r=-\rho g$. The Timmes EOS is used to re-compute the temperature and entropy. 
\end{itemize}
The thermodynamic variables obtained using the method above are extremely similar to the 1D output ones ($<1\%$ difference). Together with the composition profiles, they are saved in the ``imodel'' file that we include as Supplementary material. These are used as initial conditions for the 3D simulations.

\begin{table*}
\centering
\footnotesize
\caption{Properties of the three high-resolution 3D hydrodynamics neon-shell simulations presented in this study: boosting factor of the driving luminosity $\epsilon$; simulated physical time $\tau_\text{sim}$; average global root-mean-square convective velocity $v_\text{rms}$; convective turnover time $\tau_\text{c}$; time spent in quasi-steady state $\tau_\text{q}$; number of convective turnovers simulated in quasi-steady state $n_\text{c}$; average upper boundary entrainment velocity $v_\text{e}^\text{up}$; average lower boundary entrainment velocity $v_\text{e}^\text{low}$; average upper boundary bulk Richardson number $Ri_\text{B}^\text{up}$; average lower boundary bulk Richardson number $Ri_\text{B}^\text{low}$; number of CPU core-hours required to run the simulation.}\label{tab:1}
\begin{tabular}{ccccccccccc}
\hline
\hline
$\epsilon$&$\tau_\text{sim}$ (s)&$v_\text{rms}$ (cm s$^{-1}$)&$\tau_\text{c}$ (s)&$\tau_\text{q}$ (s)&$n_\text{c}$&$v_\text{e}^\text{up}$ (cm s$^{-1}$)&$v_\text{e}^\text{low}$ (cm s$^{-1}$)&$Ri_\text{B}^\text{up}$&$Ri_\text{B}^\text{low}$&CPU-core hrs ($10^6$ hr) \\
\hline
1&3182&7.04 $\times$ 10$^5$&100&700&7&1.36 $\times$ 10$^3$&1.17 $\times$ 10$^2$&493&2214&14.0\\
10&1004&1.54 $\times$ 10$^6$&46&700&15&6.97 $\times$ 10$^3$&6.16 $\times$ 10$^2$&104&492&4.9\\
100&291&3.53 $\times$ 10$^6$&23&150&6&7.56 $\times$ 10$^4$&5.63 $\times$ 10$^3$&19&103&1.4\\
\hline
\end{tabular}
\end{table*}

\subsection{Setup for 3D PROMPI simulations}
We chose to model the neon-burning shell because it is an environment that had not been previously studied with hydrodynamics simulations, and because it extends the library of burning shells modelled with the \texttt{PROMPI} code \citep{2007ApJ...667..448M}. In this way, we can compare results for the neon shell to the ones for carbon \citep{2019MNRAS.484.4645C} and oxygen \citep{2007ApJ...667..448M} shells studied with the same code, as well as other works with different hydrodynamics codes. The recent code comparison study of \citet{2022A&A...659A.193A} shows that \texttt{PROMPI} is fully consistent with other hydrodynamics codes, increasing the confidence in our new findings. The simulations in this study go beyond the current state-of-the-art in several aspects. First, the initial conditions come from a 1D stellar evolution model employing the current state-of-the-art input physics from widely used large grids of 1D models, as described above in detail. Second, convection is driven using an explicit nuclear network for neon burning including the $^4$He\footnote{The $^4$He abundance is assumed to be at nuclear equilibrium, which is reasonable for this late burning stage.}, $^{16}$O, $^{20}$Ne, $^{24}$Mg and $^{28}$Si isotopes, the reactions that link them, and the corresponding energy generation. Third and most importantly, our set of simulations includes a nominal luminosity run, in which the energy generation and reaction rates are the ones used in the 1D input model, so no additional difference from the input model is introduced.  These three aspects and the high local resolution represent an unprecedented level of realism. 
It also means that the 3D simulation obtained with the nominal luminosity provides conclusions directly applicable to 1D models, without needing to extrapolate as is usually necessary for going back to 1D \citep[e.g.][]{2019MNRAS.484.4645C}. \\
Following the “box-in-a-star” approach \citep{2016RPPh...79j2901A}, we built around the neon convective shell a cube with a side length of 0.64 $\times$ 10$^8$ cm, and a resolution of 512$^3$ cells. This is consistent with studies \citep[e.g.][]{2017MNRAS.471..279C,2022A&A...659A.193A} who tested different resolutions for hydrodynamics models and found that entrainment rate measurements are already converged below a resolution of 512$^3$. For the model domain boundaries, we assume reflective boundary conditions in the vertical directions and periodic boundary conditions in the horizontal directions. In addition, for the lower domain boundary we include a damping region between $3.40 < r < 3.46 \times 10^8$ cm in order to mimic the downward propagation of low-speed gravity waves, according to the method described in \citet{2017MNRAS.471..279C}.

\subsection{Entrainment rate determination}
The goal of the present work is to investigate the presence of turbulent entrainment in realistic 3D simulations of stellar interiors, to determine its strength and impact on the stellar structure, and to compare it to previous results. To obtain an adequate number of data points to constrain the entrainment law parameters, we completed a set of simulations from the same initial conditions (described above), each with a different boosting of the driving luminosity (a similar approach was used for the carbon shell in \citealt{2017MNRAS.471..279C}). The boosting of the luminosity is achieved here by multiplying the nuclear energy generation rate predicted from the 1D model by an artificial boosting factor of 1 (nominal case, non-boosted), 10 or 100. A larger boosting factor enhances the convective velocities, which are driven by the nuclear reactions, so turbulence is more vigorous. We list in Table \ref{tab:1} the most important properties of our three neon-shell simulations. The convective turnover time is defined as the time that the flow takes to cover twice the radial extension of the convective zone. The quasi-steady state is established in the simulations after the initial transient phase, during which convection develops throughout the original extent of the convective zone in the 1D model. As it can be inferred from the Table, increasing the boosting factor results in larger convective velocities and shorter convective turnover times. The simulations have different time ranges, depending on their luminosity and convective velocity, but all of them were run for long enough to have statistical significance (more than 5 convective turnovers). The nominal-luminosity run is the longest, due to its slow convective velocities, and more than 3000 seconds of evolution were simulated (using 14 million CPU-core hours).\\
The boundaries of the convective zone were defined using the jumps in chemical composition of the layers, as done in previous works \citep[see][]{2017MNRAS.471..279C,2019MNRAS.484.4645C}. Considering the boundary width and the mesh size, we note that the boundaries are resolved in our simulations with a number of radial grid points from $\sim$10 (lower boundary, nominal luminosity) to 60 (upper boundary, largest boosting), keeping in mind that the upper boundary is wider than the lower one, given its larger penetrability, and that a larger boosting factor increases the speed of the fluid, which smooths the boundaries and extends their width. Our boundary resolution is close to the one used in the boundary analysis of \citet{2019MNRAS.484.4645C}. An example is shown in Fig.\,\ref{fig:4} for the upper boundary of the nominal-luminosity run. \\
The entrainment velocity was obtained by averaging the time derivative of the boundary location over the largest possible time window. The final values for the entrainment velocity, for both the upper and lower boundaries of each simulation, can be found in Table \ref{tab:1}. A larger boosting factor increases convective velocities and in particular the convective boundary mixing, so a stronger entrainment is expected. Also, the entrainment velocity is always smaller for the lower boundary than for the upper one, due to the larger stiffness of the lower boundary itself. \\
The entrainment rate (i.e. entrainment velocity $v_\text{e}$ divided by convective velocity $v_\text{rms}$), that we can express according to \citet{2007ApJ...667..448M}:
\begin{equation}
E = v_\text{e}/v_\text{rms} = A\ Ri_\text{B}^{-n}
\end{equation}
depends on the “bulk Richardson number” $Ri_\text{B}$, which can be seen as a measure of the convective boundary stiffness. We refer to \citet{2019MNRAS.484.4645C} for definition and computation of the bulk Richardson number, having adopted here the same criteria. In particular, for integrating the squared Brunt-V{\"a}is{\"a}l{\"a} frequency $N^2$ we have set the length scale of turbulent motions across the boundary equal to $1⁄12$ of the local pressure scale height. This particular definition was chosen because it represents in our simulations the range that encloses completely the peak in $N^2$ at the convective boundary. The resulting values of the bulk Richardson numbers for each simulation, averaged over the same time window as the entrainment rate, can be seen in columns 9 and 10 of Table \ref{tab:1}. As expected from theory, the bulk Richardson number decreases with increasing the boosting factor, since convection and the root-mean-square velocity become stronger. This explains the negative exponent in the entrainment law. Also, the bulk Richardson number is always $\sim$5 times larger for the lower boundary than for the upper one, and this is explained by the larger stiffness of the lower boundary. With three boosting factors, we obtain a total of six measurements to estimate the parameters of the entrainment law.

\begin{figure}
\centering
\footnotesize
\includegraphics[trim={0cm 0 0cm 0cm},clip,width=0.5\textwidth]{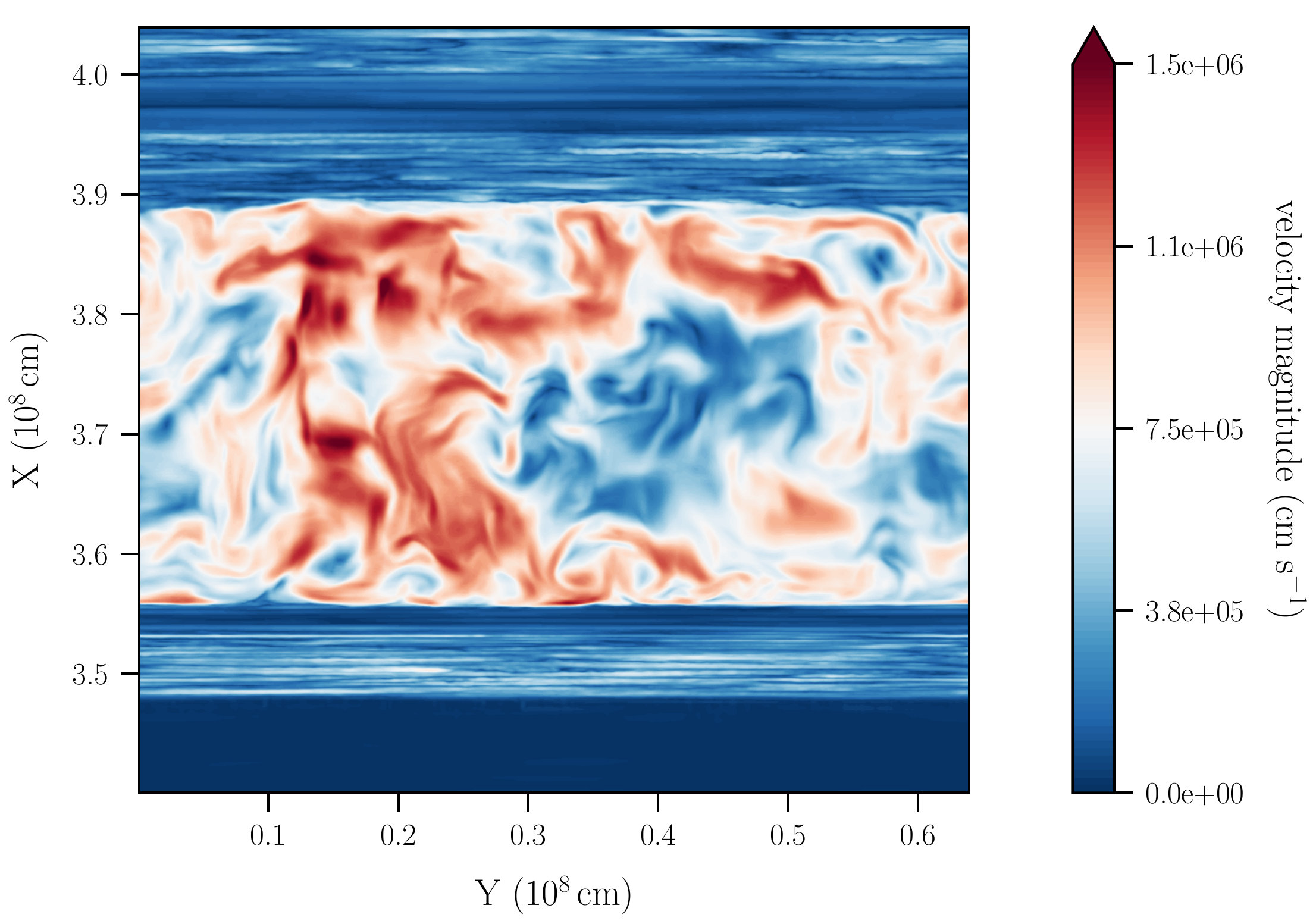}
\caption{Vertical cross section of the velocity magnitude (values in colour scale) at the end of the nominal-luminosity simulation. The high-velocity eddies between a radius of about 3.55 and 3.9 $\times$ 10$^8$ cm clearly identify the convective zone. A complete video for the evolution of the 10-times boosted simulation is available online as Supplementary material.}\label{fig:1}
\end{figure}
\begin{figure}
\centering
\footnotesize
\includegraphics[trim={0cm 0 0cm 0cm},clip,width=0.5\textwidth]{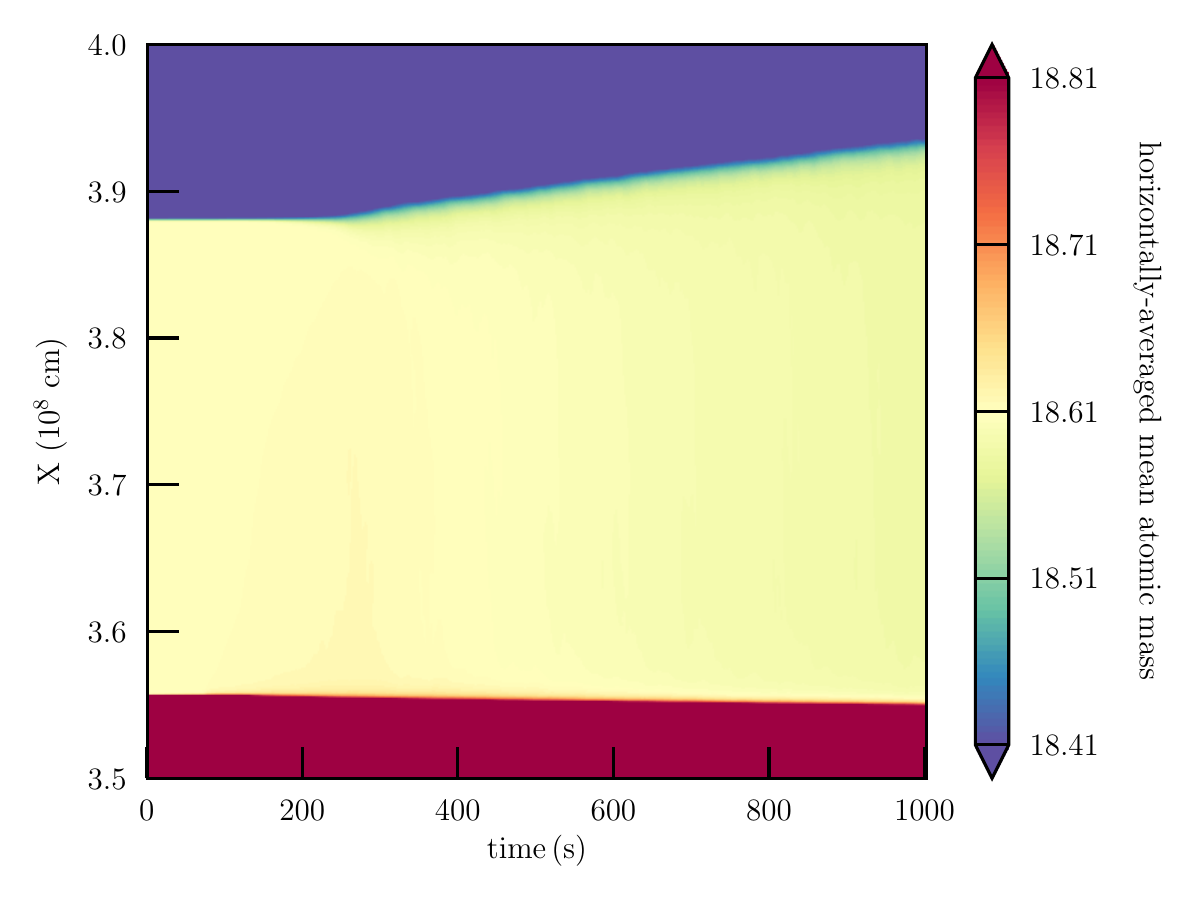}
\caption{Time evolution of the horizontally-averaged mean atomic mass for the 10-times boosted simulation. The convective zone (in yellow), after an initial transient of about 250 s, grows due to entrainment.}\label{fig:2}
\end{figure}

\section{Results}
\begin{figure}
\centering
\footnotesize
\includegraphics[trim={0cm 0 0cm 0cm},clip,width=0.48\textwidth]{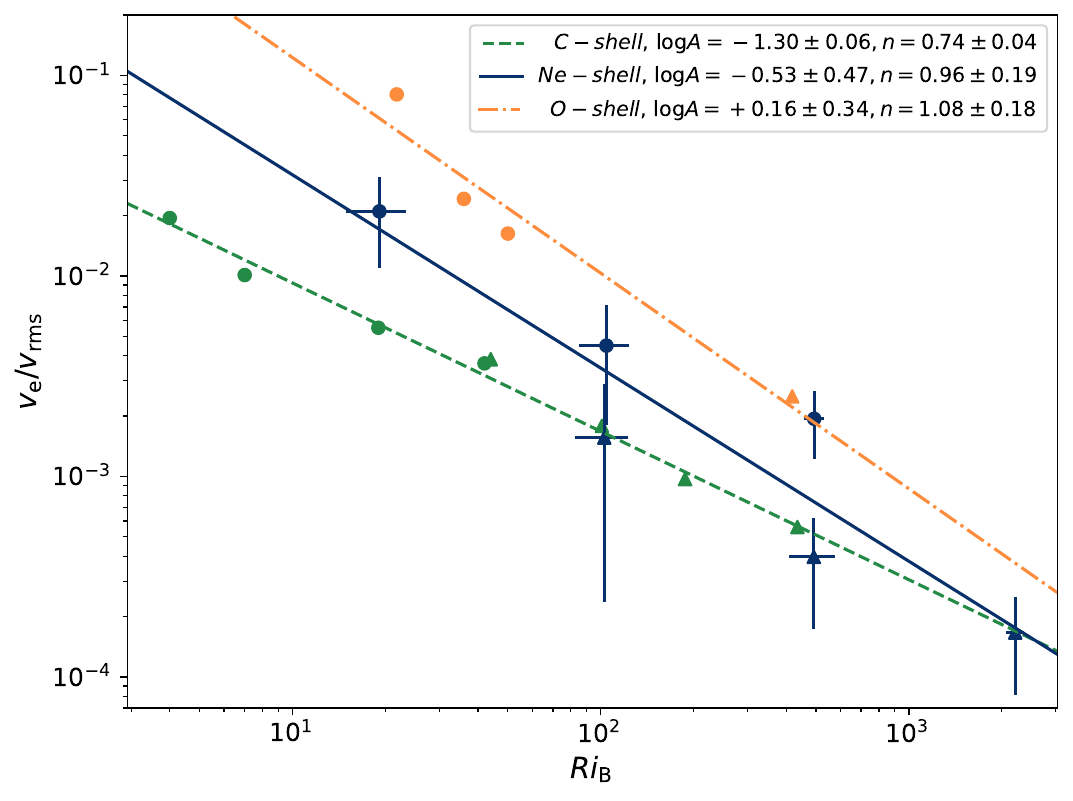}
\caption{Entrainment rate versus bulk Richardson number in log scale, measurements from \texttt{PROMPI} simulations and respective linear regressions: Ne-shell from this study (in blue, solid), C-shell from \citet{2019MNRAS.484.4645C} (in green, dashed), O-shell from \citet{2007ApJ...667..448M} (in orange, dot-dashed). Triangles are measurements for the lower convective boundary, circles are for the upper boundary. Error bars (only for Ne-shell) are standard deviations. Parameter estimates for the entrainment law $v_\text{e}/v_\text{rms} = A\ Ri_\text{B}^{-n}$ are listed in the legend for each study.}\label{fig:3}
\end{figure}

\begin{figure}
\centering
\footnotesize
\includegraphics[trim={0cm 0 0cm 0cm},clip,width=0.5\textwidth]{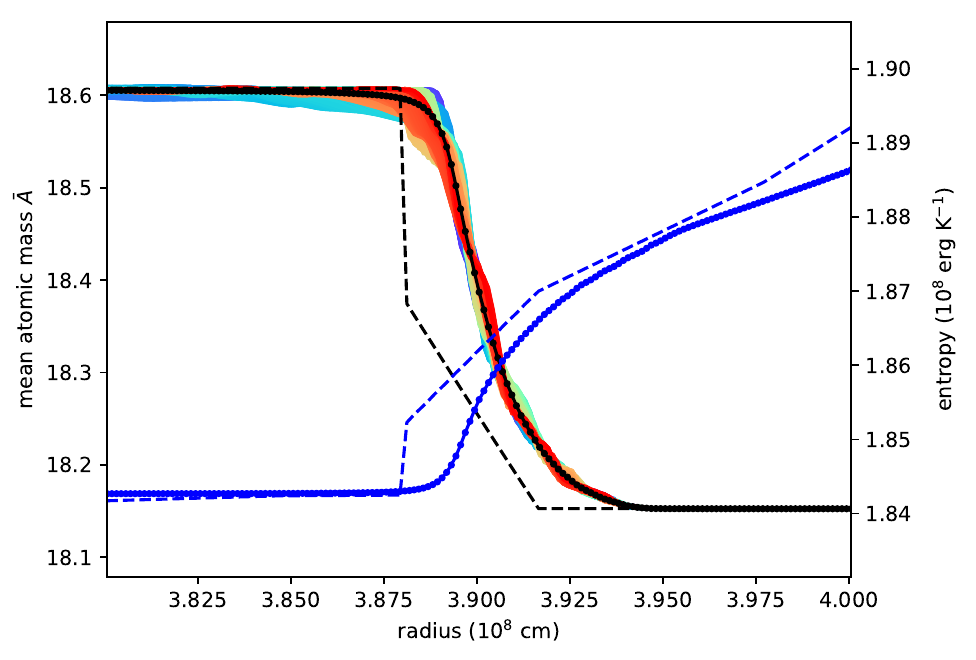}
\caption{Radial profiles of the mean atomic mass $\Bar{A}$ (black) and entropy (blue). Dashed lines are the 1D input model. Solid lines are the horizontal average of the profiles at the end of the nominal-luminosity 3D simulation, with the mesh grid explicitly indicated by dots. In colour, 512 individual (non-averaged) profiles of $\Bar{A}$ at the end of the nominal-luminosity 3D simulation. These profiles reveal important flaws in the two most common CBM prescriptions used in 1D stellar models: boundaries are not step functions as assumed in step overshooting, and both entropy and chemical composition are mixed with entrainment, not just the chemical composition as assumed in diffusive CBM.}\label{fig:4}
\end{figure}

In Fig.\,\ref{fig:1} we show a cross section of the velocity magnitude at the last timestep in the nominal-luminosity neon-shell model. Other simulations have higher velocity magnitude but very similar flow patterns. It is easy to recognize the central convective zone, characterized by eddies and plumes with high speed, in contrast to the low-speed horizontal gravity waves in the upper and lower stable regions. Furthermore, we can see in Fig.\,\ref{fig:2} the time evolution of the horizontally-averaged mean atomic mass for the 10-times boosted-luminosity simulation of the neon shell. The convective zone (in yellow), growing due to entrainment after an initial transient, can be clearly distinguished from the upper and lower stable regions using the chemical composition. 
\\Having extracted the entrainment rates and bulk Richardson numbers from the three simulations (see Table \ref{tab:1}), we are now able to estimate the parameters of the entrainment law. The line of best fit in log scale for the present neon-shell simulations is shown in Fig.\,\ref{fig:3}, alongside \texttt{PROMPI} simulations for the carbon \citep{2019MNRAS.484.4645C} and oxygen \citep{2007ApJ...667..448M} shells. Here, we also conducted a statistical study on the values that the variables assume at each timestep of our simulations, treating them as single instantaneous measurements. In this way, we included in the plot the error bars for neon measurements corresponding to the standard deviations for both the bulk Richardson number and the entrainment rate. \\
Estimates of the entrainment law parameters $A$ and $n$ are listed in the legend of Fig.\,\ref{fig:3} for the three burning stages studied to date: carbon, neon and oxygen. Obtaining parameters from different burning stages helps determine whether or not entrainment varies during the evolution of massive stars. Fig.\,\ref{fig:3} shows that the entrainment parameters obtained for the neon shell fall in between the ones for carbon and oxygen, indicating that entrainment takes place in a similar way in multiple burning stages in massive stars. This finding is also compatible with previous hydrodynamics simulations of convective zones in stellar interiors \citep{2009A&A...501..659M,2013ApJ...773..137G,2016ApJ...833..124M,2021A&A...653A..55H}. The errors estimated for the parameters do not always make them compatible between different shells, suggesting that these uncertainties are possibly underestimated. The large uncertainties and the lack of data for more burning shells do not allow us yet to confirm whether there is an evolution of the entrainment parameters with the progression of the burning phases or not.
\\Most importantly, our new simulations establish the occurrence of strong turbulent entrainment in realistic stellar conditions found in 1D models (carbon shell simulations used large boosting factors $\ge$1000 and oxygen shell simulations used an unusual initial 1D structure, as discussed above). The best fit for our neon simulations is $n=0.96\pm0.19$, which is compatible with the value of 1 expected from geophysical studies \citep{1991AnRFM..23..455F}. Estimates for parameter $A$ are more uncertain, considering the fitting done in log scale and the large dispersion of the measurements, but our neon-shell results indicate that values likely lie between 0.1 and 1.0. Furthermore, our 3D simulations also reveal important flaws in the two CBM prescriptions commonly used for 1D: (1) boundaries are not step functions as assumed in the overshooting prescriptions \citep{1991A&A...252..179Z,2012A&A...537A.146E} and (2) both entropy and chemical composition are mixed with entrainment, not just the chemical composition as assumed in the diffusive CBM prescriptions \citep{1996A&A...313..497F,2000A&A...360..952H} (see Fig.\,\ref{fig:4}). Interestingly, these 3D CBM features are compatible with the latest asteroseismic studies \citep{2021NatAs...5..715P}. It is therefore crucial to introduce entrainment into 1D stellar models. So far, only the two studies of \citet{2013ARep...57..380S} and \citet{2021MNRAS.503.4208S} have explored the implementation of turbulent entrainment, as found in our realistic 3D hydrodynamics simulations, into 1D models. They tested entrainment for CBM in the convective core of main sequence stars employing the entrainment law from geophysics \citep{1991AnRFM..23..455F} applied to stellar environments \citep{2007ApJ...667..448M}. These works mainly used $n=1$ and the parameter $A$ calibrated to asteroseismic measurements \citep{2013ARep...57..380S} and observed main sequence width \citep{2021MNRAS.503.4208S}. They found that employing the entrainment law for CBM in 1D models, with carefully chosen $A$ values around $10^{-4}$, was able to reproduce the evolution and observational features of stars in the main sequence. On the other hand, fitting the law with hydrodynamics models returns $A$ values not smaller that $\sim 0.01-0.1$ \citep[see][]{2013ApJ...773..137G,2016ApJ...833..124M,2021A&A...653A..55H}. This strongly motivates using the entrainment law also for late-phase convective shells in massive stars, even though to date no 1D stellar model has explored this possibility. 
\\The results we present here have been obtained within the context of advanced stellar phases. The question arises whether these findings can be equally applied to earlier stages. If the regime of CBM is different (e.g.\ radiative diffusion instead of neutrino cooling), there is no guarantee that the same parametrization can be applied (see \citealt{2015A&A...580A..61V} for a treatment of CBM regimes). Nevertheless, we still expect significant mixing of both entropy and composition in the CBM region (beyond the Schwarzschild/Ledoux boundary). This important question motivates the need for running more hydrodynamics simulations of the earlier stages in massive stars.


\section{Discussion and Conclusions}
With this study, we present the most realistic 3D simulations of convection in massive stars to date. Significant entrainment is found to be present when using the exact conditions from current 1D stellar models. Combined with the promising results of using entrainment in main-sequence 1D models \citep{2013ARep...57..380S, 2021MNRAS.503.4208S}, the road ahead is exciting for stellar evolution and the possibility of finally going beyond the decades-old mixing-length theory \citep{1958ZA.....46..108B}. This being said, it will still be a long and arduous journey. Our box-in-a-star simulations represent an essential step towards global simulations also including rotation and magnetic fields, though we do not expect these additional processes to affect the driving of entrainment and change our main conclusions. Furthermore, the $A$ values found in our realistic 3D simulations of the advanced phases in massive stars are several orders of magnitude larger than those found using calibration to observations on the main sequence, although the same entrainment law can in principle apply to all phases. There are several possible reasons for this large discrepancy in $A$ values inferred from 3D and observational constraints. Entrainment may only be very vigorous in the advanced phases, characterized by higher Mach-number flows and by neutrino-dominated energy transport and losses. Furthermore, starting from the wrong 1D model may lead to an overestimated entrainment rate. To solve this problem, on the one hand running the 3D hydrodynamics simulations for a longer time range will reduce the dependence on the possibly inaccurate initial conditions and predict more precise entrainment law $A$ values. Recent studies \citep{2020MNRAS.491..972A, 2022A&A...659A.193A} already show first signs of the mass entrainment rate slowing down with time, possibly approaching an equilibrium state. Running very long hydrodynamics simulations is still computationally prohibitive, but the rapid progress of parallelizing techniques and graphics processing units (GPUs) will make these simulations feasible in the coming years. On the other hand, multi-D models have shown that initial conditions from 1D stellar evolution models are inaccurate. Employing the entrainment law in 1D, especially for late convective phases when the Mach number is high and the law is expected to be more precise, will ensure a more realistic modelling of CBM. With future 1D models including entrainment and providing more realistic initial conditions for 3D simulations, we expect an improved convergence of results between 1D and 3D. \\
With concerted efforts in different disciplines such as stellar modelling, hydrodynamics simulations and asteroseismic observations, we expect the convergence between 1D and 3D model results and observations to provide new understanding of the structure and evolution of stars. This will impact many sub-fields in Astrophysics, such as supernova explosion studies. Producing accurate progenitor models, in particular predictions of the core masses, will cast new light on the origins and details of the explosion mechanism, the success or failure of the explosion and the nature of the remnant (neutron star or black hole). At minimum, we expect the next generation of 1D models including entrainment to significantly affect the final-initial mass relation for stars of all masses. Indeed, changes in predicted core masses up to 70\% \citep{2020MNRAS.496.1967K, 2021MNRAS.503.4208S} will strongly modify mass limits for the production of different remnants (white dwarfs, neutron stars and black holes) and possibly solve long-standing puzzles such as the red supergiant problem \citep{2009ARA&A..47...63S}. In addition, deviations from spherical symmetry from this and future 3D simulations will improve initial conditions for supernova simulations \citep{2015MNRAS.448.2141M, 2019ApJ...881...16Y}. The synergy between improved stellar models and recent measurements from gravitational waves will also contribute to tackle open problems such as the black hole mass gap. In this context, the impact is expected to be even larger in cases where multiple convective burning shells interact or merge, which resets the internal structure \citep[density and composition,][]{2009ApJ...702.1068T, 2020MNRAS.494L..53F, 2021MNRAS.500.2685C} and explodability \citep[e.g. core compactness,][]{2011ApJ...730...70O, 2016ApJ...818..124E} of massive stars. Including entrainment in 1D models will thus likely redefine the extent of the black hole mass gap.

\section*{Acknowledgements}
RH acknowledges support from the World Premier International Research Centre Initiative (WPI Initiative), MEXT, Japan and the IReNA AccelNet Network of Networks (National Science Foundation, Grant No. OISE-1927130). CG has received funding from the European Research Council (ERC) under the European Union’s Horizon 2020 research and innovation program (Grant No. 833925). WDA acknowledges support from the Theoretical Astrophysics Program (TAP) at the University of Arizona and Steward Observatory. CG, RH, and CM acknowledge ISSI, Bern, for its support in organising collaboration. This article is based upon work from the ChETEC COST Action (CA16117) and the European Union’s Horizon 2020 research and innovation programme (ChETEC-INFRA, Grant No. 101008324). The authors acknowledge PRACE for awarding access to the resource MareNostrum 4 at Barcelona Supercomputing Center, Spain, and the STFC DiRAC HPC Facility at Durham University, UK (Grants ST/P002293/1, ST/R002371/1, ST/R000832/1, ST/K00042X/1, ST/H008519/1, ST/ K00087X/1, ST/K003267/1). The University of Edinburgh is a charitable body, registered in Scotland, with Registration No. SC005336. For the purpose of open access, the author has applied a Creative Commons Attribution (CC BY) licence to any Author Accepted Manuscript version arising.

\section*{Data availability}
The data underlying this article will be shared on reasonable request to the corresponding author.




\bibliographystyle{mnras}
\bibliography{article} 


\bsp	
\label{lastpage}
\end{document}